\documentclass[reprint,prd,aps,superscriptaddress,altaffilletter,amssymb,showpacs,nofootinbib,twocolumn,showkeys,a4paper]{revtex4}
\usepackage[dvips]{graphicx} 
\usepackage[usenames]{color}
\usepackage[latin1]{inputenc}
\usepackage[english]{babel}
\usepackage{multirow}
\usepackage{verbatim}
\usepackage{amsmath,amssymb}
\usepackage{rotating}

\usepackage{subfigure}

\newcommand{\be}{\begin{equation}}
\newcommand{\ee}{\end{equation}}
\newcommand{\bea}{\begin{eqnarray}}
\newcommand{\eea}{\end{eqnarray}}

\newcommand{\ben}{\begin{eqnarray}}
\newcommand{\een}{\end{eqnarray}}

\newcommand{\neff}{{N_{\mathrm{eff}}}}
\newcommand{{\ngwcs}}{{N_{\mathrm{GW}}^{\mathrm{CS}}}}
\newcommand{{\nextra}}{{N_{\mathrm{extra}}}}
\newcommand{\fd}{\ensuremath{f_{10}}}
\newcommand{\ogw}{\Omega_{\rm {gw}} h^2}

\newcommand{\ns}{\ensuremath{n_{\mathrm{s}}}}
\newcommand{\Ob}{\ensuremath{\Omega_{\mathrm b}}}
\newcommand{\Oc}{\ensuremath{\Omega_{\mathrm c}}}

\newcommand{\Obhh}{\ensuremath{\Ob h^{2}}}

\newcommand{\Ochh}{\ensuremath{\Oc h^{2}}}

\newcommand{\optdepth}{\kappa}
\newcommand{\As}{A_\mathrm{s}}

\begin{document}
\title{Correlations between cosmic strings and extra relativistic species}

\author{Joanes Lizarraga}
\email{joanes.lizarraga@ehu.es}
\affiliation{Department of Theoretical Physics, University of the Basque Country UPV/EHU, 48080 Bilbao, Spain}

\author{Irene Sendra}
\email{irene.sendra@ehu.es}
\affiliation{Department of Theoretical Physics, University of the Basque Country UPV/EHU, 48080 Bilbao, Spain}

\author{Jon Urrestilla}
\email{jon.urrestilla@ehu.es}
\affiliation{Department of Theoretical Physics, University of the Basque Country UPV/EHU, 48080 Bilbao, Spain}

\date{26/07/2012}

\begin{abstract}

The recent observation that the Cosmic Microwave Background (CMB) may prefer a neutrino excess has triggered a number of works studying this possibility. The effect obtained by the non-interacting massless neutrino excess could be mimicked by some extra radiation component in the early universe, such as a cosmological gravitational wave background. Prompted by the fact that  a possible candidate to source those gravitational waves would be cosmic strings, we perform a parameter fitting study with models which considers both  cosmic strings and the effective number of neutrinos as free parameters, using CMB and non-CMB data. We find that there is  a correlation  between cosmic strings and the number of extra relativistic species, and that strings can  account 
for  all the extra radiation necessary.  In fact, CMB data prefer strings at a $2\sigma$ level, paying the price of a higher extra radiation component. CMB data also give a moderate preference for a model with $n_s=1$. The inclusion of non-CMB data lowers both the preference for strings and for the extra relativistic species.
\end{abstract}
\maketitle

\section{Introduction}

The analysis of Cosmic Microwave Background (CMB) has been invaluable for our understanding of our universe. There is now a standard cosmological model that the community agrees upon, which fits the data very well. Most of the unanswered questions lie in the link between the cosmological standard model and the high energy standard model.  It is in the small details and small windows that the data allow us to try different approaches where a promising road to understanding the constituents of the cosmological standard model lays.

As an example of what we mean we can consider cosmic strings \cite{VilShe94,Hindmarsh:1994re}. These are interesting entities that  many of the possible high energy physics models for inflation predict to have been formed in the early universe, and their detection could give invaluable information about the physics of the early universe. The CMB anisotropy coming from strings has been constrained to be less than 10\% \cite{Wyman:2005tu,Battye:2006pk,Fraisse:2006xc,Bevis:2007gh,Urrestilla:2007sf,Battye2010,Urrestilla2011,Dvorkin2011}, and predictions on CMB signals at small angular scales \cite{Fraisse:2007nu,Pogosian:2008am} and on B mode polarization \cite{Bevis:2007qz,Pogosian:2007gi,Urrestilla:2008jv,Mukherjee:2010ve} show that new CMB measurements will further constraint models with strings.
Cosmic strings are also candidates for the generation of other observable astrophysical phenomena such as high
energy cosmic rays, gamma ray burst and gravitational waves \cite{VilShe94,Hindmarsh:1994re, khlopov1, khlopov2}.

Another example could be the observation that several recent analysis on cosmological data seem to indicate: it has been showed \cite{Keisler2011, Dunkley2011} that  the effective number of neutrinos  ($\neff$) may be higher  than what one would expect from only three different neutrino species. This could be interpreted as more than three neutrino species being present in the early universe, which seems to contradict what we know from high energy physics. But it  can also be understood as the effect of  having some extra radiation component in the early universe. 

Along these lines, some groups have been investigating the option of gravitational waves being the extra component \cite{Smith2006}. Those gravitational waves could account for the extra radiation component needed to explain the observed  $\neff$. The Cosmic Gravitational Wave Background (CGWB) behaves as non-interacting massless particles \cite{Misner1974}, so their effect on the CMB is the same as massless neutrinos when their energy-density perturbations are adiabatic, as one would expect from inflation.  In other scenarios, where the perturbations on the CGWB energy-density are non-adiabatic the effects on the CMB from CGWB will differ from those produced by massless neutrinos \cite{Sendra:2012wh}.

One important question, though, is to give the source of those gravitational waves. If they were there at the early universe, what was creating them? One of the possible answers put forward in the literature has been that the aforementioned cosmic strings could be the source of this CGWB \cite{Sendra:2012wh}. 
Besides CMB anisotropies, cosmic strings would also be (active) sources of gravitational waves. There have been several works trying to obtain the GW spectra coming from cosmic strings, although there is still some debate in the community \cite{Damour2000,Damour2001,Siemens:2001dx,Siemens:2003ra,Siemens2006,Siemens2007,Olmez2010}. A full field theory simulation  would be very helpful. The loop distribution of strings is not completely understood either \cite{Vanchurin:2005pa,Ringeval:2005kr,Olum:2006ix,Polchinski:2006ee,Dubath:2007mf,Polchinski:2007rg,Rocha:2007ni,Lorenz:2010sm,Hindmarsh:2008dw,BlancoPillado:2010sy,BlancoPillado:2011dq}, and that parameter could be key in estimating the gravitational wave production from strings. Besides, the decay products of strings are also important, since strings could be decaying directly into gravitational modes, or could decay into other particles. For example, pulsar timing bounds on cosmic string tension can change significantly due to those factors mentioned above,
 {\it i.e.}, loop production and decay mechanisms.

All in all, the excess energy stored in relativistic components could come from gravitational waves originated in cosmic strings. In \cite{Sendra:2012wh} the authors discussed this possibility and gave an upper bound on the cosmic string tension $G{\mu} \leq 2\times 10^{-7}$ at $95\%$ confidence
level (C.L.). However, that zero order  approach can be improved by noting the following: if the excess in  $\neff$ comes from cosmic strings, cosmic strings were there in the early universe. Then, cosmic strings would have also contributed to the temperature (and polarization) CMB anisotropies, and thus the parameter estimation should be done including the contribution from string from the beginning. 

It is known that cosmic strings have degeneracies and correlations with other parameters in the $\Lambda$CDM model. For example, it was shown in \cite{Battye:2006pk,Bevis:2007gh} that an $n_s=1$ was possible with the (then) current data if strings were included in the analysis, due to a suit of degeneracies.  Thus, it is interesting to investigate whether such correlations exist between the string contribution and  $\neff$, and what implications this may have both in the string tension and the extra neutrino species.

This is the task we undertake in this paper: we perform a Markov Chain Monte Carlo type analysis to estimate cosmological parameters, including cosmic string contributions, letting  $\neff$ be a free parameter. We use CMB data  both for relatively small $\ell$ (WMAP7 \cite{Larson2011}) and for larger $\ell$  (South Pole Telescope (SPT) data  \cite{Keisler2011}). We will also consider non-CMB experimental data given by the Hubble Space Telescope (HST) measurement of the Hubble constant ($H_0$)  by Riess {\it et al} \cite{Riess2009} and the Baryon Acoustic Oscillation (BAO) data  by Percival {\it et al} \cite{Percival2010}.
In section \ref{method} we describe the basic ingredients for our analysis, then we present our results in section \ref{results}, and then conclude in \ref{conclude}.

\section{Methodology}
\label{method}
In this section we discuss what  information the effective number of  neutrinos  parameter  $\neff$ carries, then introduce the cosmic string template, and lastly, we show  how the data analysis is done.

\subsection{Effective number of neutrinos}
\label{enn}

The amplitude and the position of the acoustic peaks of the CMB power spectrum are determined by the energy content of the
Universe before recombination, i.e., by the matter and radiation content. So in this context it is crucial to have a refined
knowledge of each of the components that could contribute to the radiation energy density at that epoch. The main component are photons, $\Omega_{\gamma}$, but other relativistic species like neutrinos can contribute to the radiation in the
form:

\be
\Omega_{rad}=\left[1+\frac{7}{8}\left(\frac{4}{11}\right)^{4/3}N_{\mathrm{eff}}\right]\Omega_\gamma\,,
\ee
where any extra radiation component is encapsulated in  $\neff$, usually known as the effective number of neutrinos. One  of its main effects is the change of the matter-radiation equality redshift, $z_{eq}$, which increases (decreases) as $N_{\mathrm{eff}}$ gets smaller (bigger). The same effect occurs if the energy density of matter,
 $\Omega_mh^2$, increases (decreases). The parameter $N_{\mathrm{eff}}$ can be expressed as  \cite{Komatsu2011}:

\be
N_{\mathrm{eff}}=3.04+7.44\left(\frac{\Omega_mh^2}{0.1308}\frac{3139}{1+z_{eq}}-1\right).
\ee
Both those quantities, $z_{eq}$ and $\Omega_mh^2$, can be obtained directly from the matter power spectrum, allowing us to constrain  $\neff$.

The high energy physics standard model predicts three neutrino flavours, giving $N_{\mathrm{eff}}=3.046$, which also accounts for some corrections
from incomplete decoupling \cite{Mangano2005}. However, analysis on  $N_{\mathrm{eff}}$ from recent cosmological data seems to prefer values  greater than $3.046$ indicating the presence of an extra relativistic energy component at the
early universe: a value of $\neff= 4.34^{+0.86}_{-0.88}$ at  $68\%$ C.L. for the combination of CMB data from WMAP7 with BAO presented in \cite{Komatsu2011} and the value of $H_0$ given  by the Hubble Space Telescope (HST) \cite{Riess2009}. Also the inclusion of CMB data at small scales, which provides tighter constraints, still shows a preference for the existence of an extra component of radiation different from neutrinos. In combination with WMAP7 data, BAO and the value of $H_0$, an $\neff=4.56\pm1.5$ at $68\%$ C.L is measured from the Atacama Cosmology Telescope  (ACT) \cite{Dunkley2011} and $\neff=3.86\pm0.42$ from the  South Pole Telescope (SPT) data \cite{Keisler2011}.

This excess of relativistic energy could be an indication of an extra neutrino, but it could also indicate that there were other types of relativistic components in the early universe:
it has been pointed out in \cite{Smith2006} that this excess can be explained with the contribution of the primordial CGWB to the total energy density of radiation.
  
As detailed in Sec.~$35.7$ of \cite{Misner1974}, it is possible to obtain an effective energy momentum 
tensor for the gravitational waves as an average of the stress energy carried by various wavelengths
\be
T_{\mu\nu}^{GW}=\frac{\langle h_{jk,\mu}h_{jk,\nu} \rangle}{32\pi},
\ee
where $h_{jk,\mu}$ represent the covariant derivative of the tensor perturbation of the metric $h_{jk}$.
For perturbations inside the horizon, gravitational waves  can be considered to be propagating in a flat, Minkowski background.  In this case,  the equation of motion for the tensor perturbation
of the metric becomes $\partial^\sigma \partial_\sigma h_{\mu \nu}=0$, whose solution is a plane-wave $h_{\mu\nu}=\mathcal{R}\{A_{\mu\nu}e^{ik_\sigma x^\sigma}\}$, with a wave vector $\mathit{k}$. We can then obtain an expression of the effective
energy-momentum tensor:
\be
T_{\mu\nu}^{GW}=\frac{A^2k_\mu k_\nu}{32\pi},
\ee
which is the same as the energy-momentum tensor of a beam of non-interacting massless particles. Thus,  the effects on the CMB and the matter power spectra of CGWB are equal to those produced by massless neutrinos. This allows us to translate any  constraint on the $\neff$ into an upper bound on the gravitational wave energy density $\ogw$, for those with a frequency larger than $\sim 10^{-15}$Hz which corresponds to the size of the sound horizon at decoupling.  We can then obtain an upper limit of the GW energy density $\ogw$, integrated over all the possible frequencies, see \cite{Maggiore2000} for further details:
\be
 \int_{0}^{^{\infty}} d\left( \ln{f}\right) h^2 \Omega_{gw}(f) = 5.6\times10^{-6}(\neff - 3.046).
\label{eq:int_Ogw}
\ee

The study of this extra radiation component is of special interest, since it carries exclusive information about the state of the primordial Universe, providing a unique window to explore the evolution of the Universe at those times. But, even if it is gravitational waves, where do they come from? One possibility would be that primordial gravitational waves (primordial tensor fluctuations) could be responsible. These are usually characterized by $r$, the tensor-to-scalar ratio. Using an upper bound on $r<0.20$ and some standard approximations \cite{Chongchitnan:2006pe}, one can see that $\ogw < 10^{-14}$, which corresponds to a tiny  $\neff$. Another (maybe more promising) source of gravitational waves could be cosmic strings, which we describe in the following section. 

\subsection{Cosmic string power spectra}
\label{csps}

Cosmic strings are one type of cosmic topological defect that could be formed in the early universe; and if formed, they would still be around us. In high-energy physics models of inflation, cosmic strings are ubiquitous, since they appear generically in models of SUSY grand unification and in brane inflationary models. 

Calculating the imprint of cosmic strings in the CMB is a difficult task, because cosmic strings  are highly non-linear objects that are active sources of the perturbations, i.e., they keep sourcing the anisotropies; unlike inflationary contributions, which are generated during decoupling and propagate freely afterwards. The thickness of cosmic strings is microscopical, though their length can be cosmological, and the time in which they evolve is also cosmological. It is thus very hard to get an analytical handle on them (though there are works on effective theories of defects, e.g., \cite{Kibble:1984hp,Bennett:1985qt,Martins:1996jp,Martins:2000cs}) and numerically is also very difficult. There are different approaches to numerically simulate cosmic strings: one can use the aforementioned effective theory (known as the "unconnected segment model", e.g., \cite{Albrecht:1997nt,Battye:2006pk,Pogosian:2008am,Pogosian:1999np,Battye:2010xz}), one can forget about the microphysical details of the core of the strings and 
treat it as an infinitely thin object (the so-called "Nambu-Goto model", e.g., \cite{Contaldi:1998mx,Landriau:2003xf,Ringeval:2005kr,Olum:2006ix,Fraisse:2007nu}), or one could try to perform a full field theoretical simulations to obtain the CMB power spectra (which is usually done using an Abelian Higgs type model, e.g. \cite{Vincent:1996qr,Vincent:1996rb,Vincent:1997cx,Bevis:2006mj,Bevis:2010gj,Moore:2001px}). All those approaches have advantages and drawbacks, but in some sense they all are complementary.  In this work, we have chosen to use the Abelian-Higgs type approach for the CMB predictions, and the string power spectra used is the one presented in \cite{Bevis:2006mj,Bevis:2010gj}.

In order to perform the parameter estimation including cosmic strings, one would need to calculate the spectra of the strings at every point in the Markov Chain. That is computationally very costly;  instead, the cosmic string spectrum is computed once for a given parameter set (see \cite{Bevis:2010gj}), and then the normalization of the spectra is left as a free parameter. This is justified in previous works, but it is clearly reasonable as the contribution from the strings is subdominant.

The normalization is given by the parameter $G\mu$, where G is Newton's constant and $\mu$ is the string tension. There is an analogous parameter that has been used in the literature previously: $\fd$. This is the fraction of the total power coming from strings at multipole $\ell=10$. Although the exact correspondence between $G\mu$ and $\fd$ is model dependent, it is worth noting that $\fd\propto(G\mu)^2$ for $\fd\ll1$.

The contribution from cosmic strings to the CGWB is even less understood than their CMB imprints.  The production of GW from strings has been studied for a number of years now \cite{Vilenkin1981,Damour2000,Damour2001,Siemens2006}, where the cusps and kinks on the strings where considered, mostly in a Nambu-Goto approximation. For those predictions, the size and distributions of loops are also of crucial importance. One of the main focus of some of those works was to study how the GW contributions from strings would change when considering cosmic superstrings, and more specifically the probability of intercommutation of the strings $p$. For the present work, we only consider solitonic cosmic strings, and will consider $p=1$, that is, the string segments always reconnect when they meet, which is an excellent approximation\footnote{This is the usual assumption, which is an approximation, albeit an extremely good one. There are however situations  in which  the probability of  intercommutation of solitonic 
strings   is not one \cite{Bettencourt:1996qe, Achucarro:2006es,Verbiest:2011kv}.}.

Even though the form of the GW spectra coming from strings is still under study (see \cite{Dufaux:2010cf} for a different approach), as is also the size and distribution of loops \cite{Hindmarsh:2008dw,Lorenz:2010sm,Shlaer:2012rj}, for the present work we will  use the approach put forward in \cite{Siemens2007,Olmez2010} since it is one of the few that gives a direct translation from $G\mu$ to gravitational waves (and hence, to $\neff$). Those authors show that under adiabatic initial conditions,  incoherent superposition of cusp bursts from a network of cosmic strings (or superstrings) can produce a spectrum of gravitational waves. Assuming a certain model, it is possible to convert any bounds on $\ogw$ into another over the string tension $G\mu$ \cite{Siemens2007, Olmez2010}; or vice versa. In \cite{Sendra:2012wh} the authors follow the procedure given in \cite{Olmez2010} which provides an analytical approximation of the CGWB spectrum produced by strings. This approximate model allows us to relate the 
constraints on the  $G\mu$ to the gravitational wave energy density (remember that the string reconnection probability $p$  will be fixed  to 1 for our study).  
In \cite{Olmez2010} two different expressions for the CGWB energy density are given depending on the string loops size. For those smaller 
than the horizon we have:
\begin{equation}
\Omega_{\rm gw}(f) \approx 5 \times 10^{-2} G \mu
\label{eq:short_spectrum}
\end{equation}
whereas the spectra for the  CGWB  energy density given  by loops of the size of the horizon is
\begin{equation}
\Omega_{\rm gw}(f) \approx 3.2\times10^{-4} \sqrt{G \mu}.  
\label{eq:long_spectrum}
\end{equation}

Once we have a value for $G\mu$ we can link it to a $\ogw$ which in turn is linked to an effective number of relativistic species $\ngwcs$ whose effect would be analogous to the one produced by the gravitational waves produced for those strings,  all according to \cite{Olmez2010}. In order to get the correspondence between $G\mu$ and $\neff$ we would need to integrate Eq.~(\ref{eq:int_Ogw}) using Eqns.~(\ref{eq:short_spectrum}) or  (\ref{eq:long_spectrum}) for sub-horizon or horizon sized loops respectively, as done in \cite{Sendra:2012wh}. The limits for the integral can be obtained approximately as follows:
the lower-bounds determine the range of validity of the approximations given in \cite{Olmez2010} where the GW spectra is almost flat and gets the main contribution from the loops in the radiation era. The upper limit is given by the horizon size at the time of the phase-transition which produced the string network.

For subhorizon string loops,  $f_{\rm min}\sim  \alpha^{-1} H_0z_{\rm eq}^{1/2}$ and $f_{\rm max}\sim \alpha^{-1} M_{\rm pl} G\mu$, with $\alpha$ some constant that drops out of our calculation, giving 
\begin{equation}
G\mu \ln\left(\frac{ G\mu M_{\rm pl} }{H_0z_{\rm eq}^{1/2}}\right) \sim \frac{5.6\times10^{-6}\ngwcs}{5\times10^{-2}h^2}\,.
\label{small}
\end{equation}
For horizon-sized loops $f_{\rm min}\sim 3.6\times10^{-18}/ G\mu\ {\rm Hz}$ and $f_{\rm max}\sim \alpha^{-1} M_{\rm pl} G\mu$, with $\alpha\sim 1$ in this case \cite{Olmez2010}, giving
\begin{equation}
\sqrt{G\mu}\ln\left(\frac{(G\mu)^2 M_{\rm pl}}{3.6\times10^{-18}\ {\rm Hz}}\right) \sim \frac{5.6\times10^{-6}\ngwcs}{3.2\times10^{-4}h^2}.
\label{large}
\end{equation}

In \cite{Sendra:2012wh}, Sendra and Smith calculated the best fit values of $\neff$ obtained in the parameter fitting to get a corresponding bound in $G\mu$ using the formulas above.  In this work we follow the reverse path: starting from the values obtained for $G\mu$ from parameter fitting, we use the formulas above to estimate the amount of gravitational waves the strings would have produced, and the effective number of relativistic species  they would correspond to.  

Note that in this work we use several different {\it effective number of relativistic species}. We  have $\neff$, which is the usual number used in the literature: this accounts for the effective number of  any relativistic species at Big Bang Nucleosynthesis, except for photons and electrons, and most commonly accounts for only the three neutrino species. As we just mentioned, the cosmic strings present in our models  will  contribute to the total relativistic species. The effective number of relativistic species analogous to the gravitational waves produced  from strings is what we call $\ngwcs$.  This contribution is another of the contributions included in $\neff$. It is necessary to stress that there is a very high level of uncertainty in $\ngwcs$; not only because it will differ between long or short loops, but mainly because the translation from $G\mu$ to $\ngwcs$ is far from understood.    Thus, our models have two different sources for the effective number of relativistic species: the three neutrinos and the strings. Therefore, a new parameter is defined: $\neff^*=\neff-3.046-\ngwcs$, which accounts for  any other extra ingredient necessary to fit the data best. A negative $\neff^*$ would tell us that either less than three neutrinos are preferred, or (more likely) that contribution from cosmic strings is too high.

\subsection{Data analysis}

\begin{table}[!htp]
\begin{tabular}{c c|c}
\hline
Parameter & Symbol & Prior \\
\hline
Baryon density & $\Omega_bh^2$ & [0.005, 0.1]           \\
Cold Dark Matter density & $\Omega_ch^2$ & [0.01, 0.99]     \\
Angular size of sound horizon &$\theta$ & [0.5, 10]       \\
Optical depth to reionization & $\kappa$ & [0.01, 0.8]       \\
Scalar spectral index & $n_s$ & [0.5, 1.5]       \\
Amplitude of scalar spectrum & $\ln(10^{10}A_s)$ &[2.7, 4]      \\
Effective number of Neutrinos & $\neff$ & [1.047, 10.0]      \\
Cosmic Strings normalization & $(G\mu)^2$ & [0, $4\times 10^{-12}$]     \\
\hline
Poisson point source power & $D_{3000}^{PS}$ & [0, 100]\\
Clustered point source power & $D_{3000}^{CL}$&[0, 100] \\
Sunyaev-Zel'dovich power & $D_{3000}^{SZ}$&[0, 100] \\
\end{tabular}
\caption{\label{prior} Priors for the parameters used in  our analysis, in the most general case with all parameters left free. For the different models we considered, some of this parameters will be kept fixed, as explained in the text.}
\end{table}

The model considered is that of a $\Lambda$CDM with the usual six parameters: Amplitude of scalar spectrum $A_s$; Baryon density $\Omega_b$; Cold Dark Matter density $\Omega_c$; Optical depth to reionization $\kappa$; Angular size of sound horizon $\theta$; and  Scalar spectral index $n_s$.
We will refer to this model as the Power Law (PL) model. These parameters are fed into {\tt CAMB} \cite{camb_notes} to get the corresponding power spectrum. 
In our analysis, one or two more parameters will be added to the PL model: 
 the effective number of neutrino species  $\neff$ (or equivalently the extra radiation component) and/or cosmic strings (parametrized as $(G\mu)^2$). As we will show later, for certain cases we find that the data prefers $n_s=1$, in a similar way to what the authors in \cite{Battye:2006pk,Bevis:2007gh,Urrestilla:2007sf,Battye:2010hg} found. Therefore, we will also consider a model with $n_s$ fixed to one. We will refer to this model as the Harrison-Zel'dovich (HZ) model. Once again, $\neff$ and $G\mu$ will be added to the HZ model.  The linear priors used for the parameters can be found in Table~\ref{prior}.

Besides the parameters mentioned above, on the following tables we also show the values of some derived parameters: the Hubble parameter $h$ ($H_0=100h$ km s$^{-1}$ Mpc$^{-1}$); the contributions of strings to the total power spectrum at $\ell=10$ $\fd$; and two extra relativistic numbers, $\ngwcs$, which is the contribution to the effective number of extra species  solely coming from strings, and $\neff^*$, which is the contribution needed once the three neutrinos and the cosmic strings contribution has been subtracted. These last two numbers are given for both, horizon sized and sub horizon sized loops.

Our models will be compared with both CMB and non-CMB  experimental data with the help of the Markov
Chain MonteCarlo (MCMC) method, using a modified version of CosmoMC \cite{Lewis:2002ah}. Specifically, the CMB  data we will use will be that of  WMAP7 \cite{Larson2011}  and SPT \cite{Keisler2011}. When data from  the SPT experiment is used, foreground contaminants have to be taken into account, therefore three extra parameters are used (normalized, as usual, at  $\ell=3000$): Poisson point source power $D_{3000}^{PS}$; Clustered point source power $D_{3000}^{CS}$; and the Sunyaev-Zel'dovich power $D_{3000}^{SZ}$. The  non-CMB experiment data that we will consider are the HST measurement of the Hubble constant ($H_0$)  by Riess {\it et al} \cite{Riess2009}
and the BAO by Percival {\it et al} \cite{Percival2010}.

\section{Results}
\label{results}

The results of fitting WMAP7 data to the PL model with  $\neff$ and/or $G\mu$ is shown in Table~\ref{wmap7}. When we only allow $\neff$ to be a free parameter and set $G\mu$ to zero, it is clear that, as it was pointed in previous works \cite{Keisler2011,Dunkley2011}, the preferred value for $\neff$ is larger than three, indicating that there is a preference for some extra radiation component in the early universe. When including cosmic strings also into the analysis,  the preferred value is found to be still larger than three, but smaller than when considering only $\neff$. Thus, not only would cosmic strings be the source for the extra radiation component by means of creating gravitational waves, but they also lower the need of the extra component; that is, $G\mu$ and $\neff$ are {\it anticorrelated}. For comparison, we give the numbers for the case where $\neff$ is fixed to 3.046, and only the string contribution is allowed to vary. 

Table~\ref{wmap7} also shows the best-fit $\chi^2$ differences between different models and data-sets, taking as the base model the one corresponding to  $PL+G\mu+\neff$ fitting for CMB data only. In other words, we compare all the other models to this one, and give the difference in $\chi^2$ in the table.  In order to compare the best-fit values, one should be careful about the number of parameters and the data-sets used to perform the likelihood analysis. For example, the first two columns of the table have the same number of parameters (7) and fit to the same data-set (WMAP7), so the $\chi^2$ can be directly compared, showing that they are similar, with a slight preference for the model with $\neff$. The third column has one more parameter (8) so the comparison is not as straightforward. A similar situation happens in the last three columns, where the number of parameters is the same in all of them (8) but the data-sets change from case to case; rendering the comparison of the goodness of fit by the best-fit likelihood value more difficult. However, note that the $\Delta\chi^2$ numbers are only a  rough tool to compare different scenarios; in order to compare different models a proper model selection analysis should be done, and moreover, models fitting different data sets should not strictly be compared this way. 

\begin{table*}[!htp]
\begin{tabular}{|c||c|c|c||c|c|c||}
\hline
Data & \multicolumn{3}{c||}{WMAP7}&   +$H_0$     & +BAO   & +$H_0$+BAO \\
\hline
Model   &   PL+G$\mu$     & PL + $\neff$   &{PL+G$\mu$+$\neff$} &    \multicolumn{3}{c||}{PL+G$\mu$+$\neff$} \\
\hline
100$\Obhh$ & 2.4$\pm$0.1 &2.21$\pm$0.06 &  2.3$\pm$0.1&  2.32$\pm$0.09 &$2.29\pm0.07$ &  2.31$\pm$0.01 \\
$\Ochh$ & 0.108$\pm$0.006 &0.18$\pm$0.03 & 0.16$\pm$0.04  & 0.118$\pm$0.02 &$0.19\pm0.03$ &  0.12$\pm$ 0.02 \\
$\theta$& 1.041$\pm$0.003 & 1.029$\pm$0.005&1.034$\pm$0.007& 1.039$\pm$0.006 &1.028$\pm$ 0.004 &  1.038$\pm$ 0.006 \\
$\optdepth$& 0.09$\pm$0.02 & 0.09$\pm$0.02& 0.09$\pm$0.02& 0.09$\pm$0.02 & $0.09\pm0.02$&  0.09$\pm$ 0.02 \\
$\ns$   &0.99$\pm$0.02  &1.00$\pm$0.02 & 1.00$\pm$0.02 & 0.99$\pm$0.02 &$1.02\pm0.02$ &  0.99$\pm$ 0.02 \\
$\ln(10^{10}\As)$& 3.10$\pm$0.05 & 3.12$\pm$0.04& 3.10$\pm$0.07& 3.12$\pm$0.06 &$3.13\pm0.05$ & 3.13$\pm$ 0.07 \\
$10^{12}(G\mu)^2$ & 0.18 ($<0.37)$ & --  & 0.15 ($<0.33$) & 0.15 ($<0.33$)  &0.10 ($<0.26$) & 0.15 ($<0.34$) \\
$\neff$ &--  & 7$\pm$2&6$\pm$2 & 3.6$\pm$0.9 &8$\pm$2 &4$\pm$1  \\
\hline
$h$     &0.74$\pm$0.04 &0.84$\pm$0.08  & 0.83$\pm$0.08& 0.74$\pm$0.02 &$0.88\pm0.07$&  0.75$\pm$ 0.02 \\
$\fd$    & 0.05 ($<0.107$) & -- & 0.04 ($<0.096$)   &0.04 ($<0.091$) & 0.03 ($<0.073$) &  0.04 ($<0.096$)\\
$\ngwcs$(sm) & 0.16$\pm$0.06 & -- & 0.18$\pm$0.08  &0.14$\pm$0.05&0.15$\pm$0.07 &  0.14$\pm$0.05\\
$\ngwcs$(lar) & 2.2$\pm$0.6 & -- & 2.8$\pm$0.8   & 2.1$\pm$0.4 &2.5$\pm$0.7 &  2.1$\pm$0.4\\
$\neff^*$(sm) & -- & 4$\pm$2 & 3$\pm$2   & 0.4$\pm$0.9& 4$\pm$2 &  0.3$\pm$0.9\\
$\neff^*$(lar) & -- & 4$\pm$2 & 0$\pm$2   & -1$\pm$1 & 1$\pm$2 &  -2$\pm$1\\
\hline
$\Delta\chi^2$ & -0.78 & -0.45 & 0 & -0.36 & -1.55 & -0.68\\
\hline
\end{tabular}
\caption{\label{wmap7} Marginalized likelihood constraints on model parameters, for three different models and with different datasets. The models differ in that $\neff$ and/or $G\mu$ are parameters that we fit for, or are fixed parameters. The first three columns corresponds to the fitting to  WMAP7 data, whereas the next three correspond to WAMP7+$H_0$, WMAP7+BAO, and WMAP7+$H_0$+BAO, respectively. The table shows the 6 usual parameters, plus the cosmic string contribution $G\mu$ and the extra radiation component $\neff$. We also give the derived parameters $h$ (the Hubble parameter) and $\fd$ (the contributions of strings to the total power spectrum at $\ell=10$). Following the discussion in Sections \ref{enn} and \ref{csps}, the values of $\ngwcs$ and $\neff^*$ are given for two cases:  subhorizon-sized string loops (sm) and  horizon-sized loops (lar). The goodness-of-fit is characterized but $\Delta\chi^2$, where the difference is taken with respect to the model in the third data column. The values shown are the means and standard deviations, whereas the $\Delta \chi^2$ shown corresponds to the case that best fits the data. In the cases where only upper limits can be placed, we give the mean value and next to it the 95\% confidence limit in parenthesis.}
\end{table*}

The value of $\neff$ is quite poorly constraint by using only  WMAP7 data \cite{Larson2011}; it is customary thus to include other type of data. As mentioned before, we use $H_0$ and BAO data, and the results can be seen in Table~\ref{wmap7}. The BAO and the $H_0$ data push most of the parameters in different ways; more specifically, BAO pushes the string contribution down and the $\neff$ up; whereas $H_0$ pushes  $\neff$ down. When considering both data sets together, the tension between the two data comes into play, but the contribution from $H_0$ is more important in that the overall value of the parameters are tilted towards the preferences of the HST value of $H_0$; maybe because  when fitting the data without the $H_0$ prior, the value of $h$ is rather high, and including the prior pushes it down considerably. Comparing the results from fitting to WMAP7 only,  to WMAP7+$H_0$+BAO, we see that the mean value for the extra neutrino species is significantly lower.

The constraints on the number of effective neutrinos gets tighter  by the inclusion of  high $\ell$ data (SPT \cite{Keisler2011}). The results can be found in Table~\ref{spt}, both with and without non-CMB data. It is noticeable that when adding the high $\ell$ data to the analysis, the effective number of neutrinos stays roughly the same or decreases.  Another difference is that the string contribution is favoured in this case: when comparing the model with only a string contribution (PL+$G\mu$), and the model with only $\neff$ (PL+$\neff$), the likelihood is better for the model with strings (note that both models have the same number of parameters). Moreover, the model that fits both strings and neutrinos gives a $2\sigma$ preference for strings.

\begin{table*}[!htp]
\begin{tabular}{|c||c|c|c|||c|c|c||}
\hline
Data & \multicolumn{3}{c|||}{WMAP7+SPT}&   +$H_0$     & +BAO   & +$H_0$+BAO \\
\hline
Model   &   PL+G$\mu$     & PL + $\neff$   & {PL+G$\mu$+$\neff$} &    \multicolumn{3}{c||}{PL+G$\mu$+$\neff$} \\
\hline
$100\Obhh$ & $2.21\pm0.05$ & 2.27$\pm$0.05 & 2.37$\pm$0.08 & $2.25\pm0.05$ & 2.40$\pm$0.07 & 2.25$\pm$0.05 \\
$\Ochh$ &  $0.108\pm0.005$ & 0.13$\pm$0.01 & 0.15$\pm$0.02 &  $0.13\pm0.01$& 0.14$\pm$  0.01& 0.13$\pm$0.01 \\
$\theta$ &  $1.040\pm0.002$ & 1.040$\pm$0.002 & 1.036$\pm$0.002 &$1.038\pm0.002$ &  1.037$\pm$0.002& 1.038$\pm$0.002  \\
$\optdepth$ & $0.08\pm0.01$ & 0.09$\pm$0.02& 0.09$\pm$0.02 &  $0.09\pm0.02$ & 0.09$\pm$0.02 & 0.08$\pm$0.01 \\
$\ns$  &  $0.96\pm0.01$ & 0.98$\pm$0.02& 1.01$\pm$0.02 &  $0.97\pm0.01$ & 1.01$\pm$0.02 &  0.98$\pm$0.01 \\
$\ln(10^{10}\As)$& $3.17\pm0.05$ & 3.14$\pm$0.07 &3.06$\pm$0.07 & $3.16\pm0.04$ &  3.01$\pm$0.06 & 3.16$\pm$0.04 \\
$10^{12}(G\mu)^2$& 0.11 ($<0.22$)   & --  & 0.24$\pm$0.09& 0.14 ($<0.27$)&0.26$\pm$0.06 & 0.14 ($<0.25$) \\
$\neff$&  -- & 3.9$\pm$0.6 & 6$\pm$1 &$3.9\pm0.5$ & 5.7$\pm$0.9 & 4.0$\pm$0.5  \\
$D_{3000}^{SZ}$&  $6\pm3$  & 6$\pm$3& 5$\pm$3 &$6\pm2$& 5$\pm$3&   6$\pm$3   \\
$D_{3000}^{PS}$&  $20\pm3$  & 20$\pm$3 & 19$\pm$3 &$20\pm3$& 21$\pm$3 & 20$\pm$ 3 \\
$D_{3000}^{CL}$&  $5\pm2$ & 5$\pm$2& 5$\pm$2  &$4\pm2$ & 6$\pm$2  & 5$\pm$2 \\
\hline
$h$     & $0.72\pm0.03$ & 0.75$\pm$0.04& 0.87$\pm$0.07 &$0.75\pm0.02$ &  0.89$\pm$0.06&  0.75$\pm$0.02  \\
$\fd$    & 0.03($<0.057$) & --  & 0.07$\pm$0.03 & 0.04 ($<0.072$)  &  0.08$\pm$0.02 & 0.04 ($<0.067$)   \\
$\ngwcs$(sm) & 0.11$\pm$0.04 & -- & 0.25$\pm$0.08   &0.14$\pm$0.04 &0.25$\pm$0.07 &  0.14$\pm$0.04\\
$\ngwcs$(lar) & 1.8$\pm$0.4 & -- & 3.4$\pm$0.7   &2.1$\pm$0.4 &3.3$\pm$0.8 &  2.2$\pm$0.3\\
$\neff^*$(sm) & -- & 0.9$\pm$0.6 & 2$\pm$1   &0.8$\pm$0.5& 2.2$\pm$0.9&  0.8$\pm$0.5\\
$\neff^*$(lar) & -- & 0.9$\pm$0.6 & -0.6$\pm$0.6  &-1.3$\pm$0.5 & -0.8$\pm$0.6&  -1.2$\pm$0.5\\ 
\hline
$\Delta\chi^2$ & -6.54 & -8.01 & 0 & -2.78 & -0.08 & -4.03\\
\hline 
\end{tabular}
\caption{\label{spt} Analogous table to Table~\ref{wmap7}, but in this case the CMB data sets used are WMAP7 and SPT data. Because of the presence of the SPT data, we have to incorporate 3 more parameters, responsible for taking care of foreground effects. }
\end{table*}

\begin{figure}[htb]
\resizebox{\columnwidth}{!}{\includegraphics{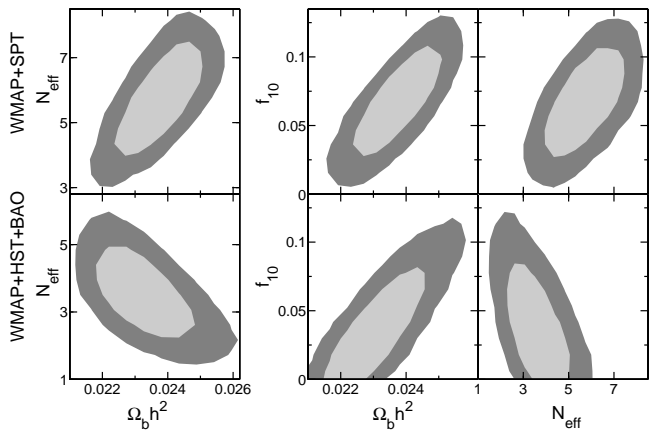}}
\caption{\label{corr} Two dimensional likelihood contours, where degeneracies between $\fd$, $\neff$ and $\Obhh$  are shown, when fitting for WMAP7+$H_0$+BAO, and when fitting for WMAP7+SPT. The shaded contours correspond to $1\sigma$ and $2\sigma$ contours. The figure shows that $\Obhh$ and $\fd$ keep the same correlation for both datasets; but $\neff$ is correlated with the other two for one case, and anticorrelated for the other.}
\end{figure}

One striking difference from the case where only WMAP7 data was used is the fact that now the string contribution and the effective number of neutrinos are {\it correlated}, that is, when the contribution of strings is pushed up, so is the effective number of neutrinos. Fig.~\ref{corr} shows the two dimensional likelihood plots for some combinations of the parameters, for two sets of data: on the one hand WMAP7+$H_0$+BAO, and on the other WMAP7+SPT. The figure shows clearly that whereas $\fd$ and $\Obhh$ are correlated in the same way for both datasets, $\neff$ is correlated or anticorrelated with the other two, depending on the datasets that are being fitted for. Thus, when fitting for the WMAP7+SPT dataset, an increase in the number of neutrinos brings an increase in the string contribution. Likewise, by allowing for strings to be present in the analysis, we not only pick up a string contribution, but the string contribution prompts the neutrino contribution to rise.

\begin{figure}[htb]
\resizebox{\columnwidth}{!}{\includegraphics{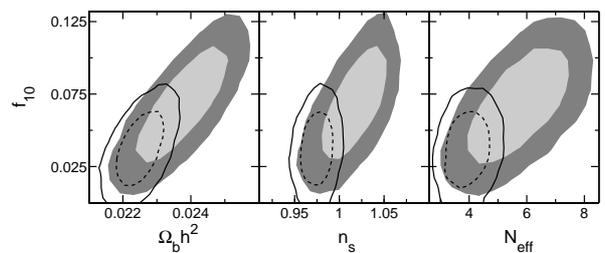}}
\caption{\label{h0bao}Two dimensional likelihood contours for some of the parameters in the model PL+$\neff$+$G\mu$. The shaded regions correspond to $1\sigma$ and $2\sigma$ confidence levels, where only CMB data are used for the analysis (i.e., WMAP7+SPT). The smaller contours depicted by  solid and dashed lines correspond also to  $1\sigma$ and $2\sigma$ confidence levels, but when also non-CMB (i.e., BAO+$H_0$) data are included.}
 \end{figure}

The inclusion of the non-CMB data, and most importantly of the HST value of $H_0$, constrains the parameters drastically better, and most of the degeneracies are broken. Fig.~\ref{h0bao} shows  two dimensional likelihood contours for some of the parameters, when fitting for WMAP7+SPT with and without non-CMB data included. The shaded lines correspond to only CMB data, whereas the solid and dashed lines take into account also the $H_0$+BAO data. It is very clear how the available parameter space has shrunk considerable, and also how  the degeneracies are broken, specially between $\neff$ and $\fd$. Note that the $2\sigma$ preference for strings is lost when non-CMB data are included.

\begin{figure}[htb]
\resizebox{\columnwidth}{!}{\includegraphics{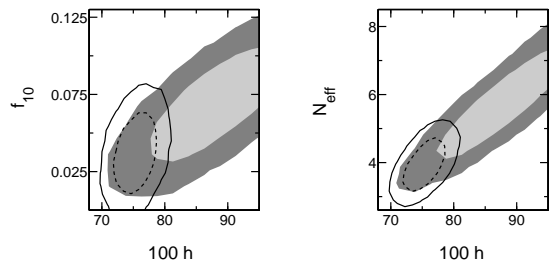}}
\caption{\label{h0baoH} Two dimensional likelihood contours for $\neff$ and $\fd$ with respect to $h$. The shaded region represents the $1\sigma$ and $2\sigma$ results obtained when only CMB data are used. Note that the mean of $h$ is higher than the value that the HST experiment gives. That is why the allowed parameter space shrinks considerable when the $h$ prior is also included (solid and dashed lines).}
 \end{figure}

The shrinking of the parameter space can be understood if one considers what is happening to the Hubble parameter. The inclusion of the HST value of $H_0$    narrows the prior space for the Hubble parameter $h\sim0.742\pm 0.036$, and  the values of $h$ that the parameter fitting prefers for the cases where the $H_0$ prior is not included are much higher. Fig.~\ref{h0baoH} shows the two dimensional likelihood contours for $h$ versus $\neff$ and $\fd$.  It is clear that the HST value of $H_0$ just catches the lower end of the contour, and it is this effect which shrinks so noticeably the parameter space.

Fig.~\ref{h0bao} also shows that high values of $n_s$ are preferred. This is  important  for inflationary models coming from hybrid supersymmetric inflation which seems to prefer values of $n_s$ larger that the ones obtained in usual analyses. In fact, $\ns=1$ lies right in the center of the confidence contours for the CMB only case, and inside the $2\sigma$ level for the other case. A similar phenomenon was studied in some previous papers \cite{Battye:2006pk,Bevis:2007gh,Urrestilla:2007sf,Battye:2010hg} where the Harrison-Zel'dovich model was then considered as a viable model to explain the data.  We have performed the same exercise, and the result is shown in Table~\ref{HZ}, where a model with $n_s=1$ (HZ) plus strings and $\neff$ is used to fit the CMB data (WMAP7+SPT). The values of the parameters change slightly from the case where $n_s$ was a free parameter, all within $1\sigma$, which is expected since $n_s=1$ was right in the middle of the confidence contours. Thus, it is not surprising that in this 
case the value of the likelihood is very similar; but bear in mind that the number of parameters is different: PL has one more parameter than HZ. If one wishes to compare models with the same number of parameters, one should compare HZ+$G\mu$+$\neff$ with PL+$G\mu$ or with PL+$\neff$; and in both those cases, the likelihood of the former is much better.  Note, however, that also in this case the value of $h$ is rather higher than what the HST value of $h$ prefers, so the inclusion of non-CMB data will disfavour this model.

\begin{table}
\begin{tabular}{|c||c|}
\hline
Data & \multicolumn{1}{c|}{SPT} \\
\hline
Model   & HZ+G$\mu$+$\neff$ \\
\hline
$100\Obhh$ & 2.31$\pm$0.04\\
$\Ochh$ & 0.14$\pm$0.01\\
$\theta$ & 1.037$\pm$0.002\\
$\optdepth$ &0.09$\pm$0.01\\
$\ns$  & 1 \\
$\ln(10^{10}\As)$ & 3.11$\pm$0.04\\
$10^{12}(G\mu)^2$& 0.19$\pm$0.07 \\
$\neff$& 5.3$\pm$0.6\\
$D_{3000}^{SZ}$& 6$\pm$2\\
$D_{3000}^{PS}$& 20$\pm$3\\
$D_{3000}^{CL}$& 5$\pm$2\\
\hline
$h$     &0.84$\pm$0.02\\
$\fd$    & 0.05$\pm$0.02 \\
$\ngwcs$(sm) & 0.20$\pm$0.04\\
$\ngwcs$(lar) & 2.9$\pm$0.4\\
$\neff^*$(sm) & 2.0$\pm$0.6\\
$\neff^*$(lar) & -0.6$\pm$0.6\\
\hline
$\Delta\chi^2$ &  -0.25\\
\hline
\end{tabular}
\caption{\label{HZ}Analogous table to Tables~\ref{wmap7} and \ref{spt}. In this case, the model we are fitting for is a HZ model, i.e., the scalar index is fixed to one ($n_s=1$). The data we are fitting for is CMB only (WMAP7+SPT).  This model fits the data better than any of the other models considered so far, even taking into account the number of parameters. Including non-CMB data would render this model less successful, partly because of the high value of $h$ that the model prefers. }
\end{table}

In all the analysis above there is one point of view that has not been taken into account: the strings that the different models prefer do in fact produce gravitational waves which will contribute to the $\neff$, but that extra contribution is not taken into account as yet. In all three tables we have included  $\ngwcs$ for subhorizon-sized and horizon-sized loops,
which is the effective number of relativistic species analogous to the gravitational waves coming from the strings.  The  numbers that we are quoting are the ones obtained following the procedure of \cite{Olmez2010,Sendra:2012wh}, and the big difference between the numbers of subhorizon-sized and of horizon-sized loops is already clear.   As we have mentioned before, this is a very rough estimate; therefore we only use it as a hint of what could be happening and the rich information we could obtain with it once the uncertainties are more under control. \\
 
In other words, the $\neff$ that we are obtaining from the parameter fitting has three possible sources: one that comes from the three species of neutrinos, another one  given by $\ngwcs$, and a third one that is not accounted for in the ingredients of our model. The parameter $\neff^*$ gives a measure of the level of those unaccounted effects. Note that the predictions for smaller sized loops need a higher level of $\neff^*$. If $\neff^*$  were exactly zero it would mean that the model is able to produce all  the $\neff$ it needs to fit the data. Note that even though the errors in these parameters are rather large,  most cases show no need for yet another contribution from another relativistic particle or source (at $2\sigma$), though do not exclude it.  The cases were  $\neff^*$ is negative  indicate that the model is not good: the amount of cosmic strings predicted is too high since their gravitational wave production is too high. 

As we have just mentioned, the case with $\neff^*=0$ is an interesting case, because it corresponds to the case where all the extra relativistic signal needed to fit the data comes from the strings, which is a possibility that most models accommodate. Therefore, we study a new set of models, where $\neff^*$ is set to zero, i.e., we allow for the possibility of having an extra radiation component, but the extra radiation comes only from the cosmic string contribution.  This way, the direct link between strings and gravitational waves is fully taken into account. The results are shown in Table~\ref{nizarra}.  There is some complicated {\it dynamics} between the parameters and the datasets. When no $h$ priors are used,  the models with horizon-sized loops allow for a larger value of $G\mu$, but with the penalty of a higher value of $h$. They fit the data rather well, and their respective $\chi^2$ are roughly the same as their base model ($PL+G\mu+\neff$), but remember that these have one parameter less. When the $h$ prior is included, the string contribution from models with horizon-sized loops is suppressed with respect to the models with subhorizon-sized loops. Nevertheless, since the $\ngwcs$ contribution from horizon-sized loops is larger than for subhorizon-sized loops for the same value of $G\mu$ (see Eqns~\ref{small} and \ref{large}), in all cases $\ngwcs$ is higher for horizon-sized loops, i.e., the extra radiation component is higher for horizon-sized loops in all cases.

\begin{table*}[!htp]
\begin{tabular}{|c||c|c|c|c||c|c|c|c||}
\hline
Data   &     \multicolumn{2}{c|} {WMAP7}  &     \multicolumn{2}{c||} {WMAP7+$H_0$+BAO}  &     \multicolumn{2}{c|} {WMAP7+SPT} &       \multicolumn{2}{c||} {WMAP7+SPT+$H_0$+BAO} \\
\hline
Loops & Subhorizon & Horizon & Subhorizon & Horizon & Subhorizon & Horizon & Subhorizon & Horizon  \\
\hline
$100\Obhh$   & 2.4$\pm$0.1 & 2.37$\pm$0.08 & 2.34$\pm$0.07 & 2.22$\pm$0.05 & 2.22$\pm$0.05 & 2.34$\pm$0.08& 2.22$\pm$0.05 & 2.20$\pm$0.05 \\

$\Ochh$      & 0.109$\pm$0.006  & 0.160$\pm$0.009 & 0.110$\pm$0.004 &  0.142$\pm$0.009 & 0.110$\pm$0.005 & 0.16$\pm$0.01 & 0.108$\pm$0.004 & 0.14$\pm$0.01 \\

$\theta$     & 1.040$\pm$0.003  & 1.032$\pm$0.002 & 1.040$\pm$0.003 & 1.033$\pm$0.003 & 1.039$\pm$0.002  & 1.036$\pm$0.001 &1.040$\pm$0.001 &1.036$\pm$0.002   \\

$\optdepth$  & 0.09$\pm$0.02 & 0.09$\pm$0.02 & 0.09$\pm$0.02 & 0.08$\pm$0.01  & 0.08$\pm$0.01 & 0.09$\pm$0.02 & 0.08$\pm$0.01 & 0.08$\pm$0.01  \\

$\ns$        & 0.99$\pm$0.02  & 1.02$\pm$0.02& 0.99$\pm$0.01 & 0.98$\pm$0.01  & 0.96$\pm$0.01 &  1.01$\pm$0.02& 0.96$\pm$0.01 & 0.98$\pm$0.01  \\

$\ln(10^{10}\As)$& 3.10$\pm$0.07 &3.08$\pm$0.06  & 3.11$\pm$0.05& 3.18$\pm$0.04 &  3.18$\pm$0.05 & 3.09$\pm$0.06& 3.16$\pm$0.04 & 3.19$\pm$0.04  \\

$10^{12}(G\mu)^2$&  0.18($<$0.37)  & 0.21 ($<$0.33)  & 0.17 ($<$0.31) & 0.06 ($<$0.13)& 0.11 ($<$0.22) & 0.18$\pm$0.08& 0.12($<$0.22) & 0.05 ($<$0.14) \\

$\ngwcs$         & 0.16$\pm$0.06 & 3.1$\pm$0.6 &  0.14$\pm$0.05 &1.6$\pm$0.4  & 0.11$\pm$0.04 & 3.1$\pm$0.7  & 0.12$\pm$0.04 & 1.5 $\pm$ 0.5 \\
$D_{3000}^{SZ}$  & - & - & - & - &5$\pm$2& 7$\pm$2&  5$\pm$2  &  8$\pm$2  \\

$D_{3000}^{PS}$  & - & - & - & - &21$\pm$3& 20$\pm$3 & 20$\pm$2 & 22$\pm$ 3 \\

$D_{3000}^{CL}$  & - & - & - & -  &5$\pm$2 & 5$\pm$2  & 5$\pm$2 &  6$\pm$2 \\
\hline

$h$              & 0.75$\pm$0.04 &0.87$\pm$0.05  & 0.74$\pm$0.02 &0.75$\pm$0.02  & 0.73$\pm$0.03 & 0.87$\pm$0.06  & 0.73$\pm$0.02 & 0.76 $\pm$ 0.02  \\

$\fd$            & 0.05($<$0.108) & 0.06($<$0.095)  & 0.05 ($<$0.088) & 0.02($<$0.037)  &  0.03 ($<$0.058) & 0.05$\pm$0.02  & 0.03($<$0.059) &0.01($<$0.037)  \\

\hline
$\Delta\chi^2$ & -0.62 & -0.04 & -0.69 & -0.77 & -5.11 & -0.66  & -5.41 & -4.96 \\
\hline
\end{tabular}
\caption{\label{nizarra} Marginalized likelihood constraints on model parameters, analogous to Table~\ref{wmap7} and \ref{spt}. In all these cases $\neff^*=0$, implying that there is a possibility of extra radiation, but all that extra radiation comes from the cosmic strings. For each different set of data, we study the cases of subhorizon-sized loops (\ref{eq:short_spectrum}) and horizon-sized loops (\ref{eq:long_spectrum}).}
\end{table*}

\section{Discussion}
\label{conclude}

In this paper we have studied the correlation between cosmic strings and extra relativistic species (or effective number of neutrino species) by fitting different models to cosmological data. We considered the possibility of cosmic strings being the source of a cosmological gravitational wave background, which in turn act as the extra relativistic species which the data seem to favor over the usual three neutrino species.  Cosmic strings are predicted in several inflationary scenarios, and therefore, they seem to be perfect candidates to seed GW and account for the extra radiation. The idea of  cosmic strings being the sources of CGWB is not new, but this work expands the previous works  
 in that the cosmic string contribution is included from the beginning, already in the parameter fitting process, and thus we are able to study different correlations 
between the different ingredients in these models. The inclusion of strings in the parameter fitting changes the value of $\neff$ that the data needs.

We have shown that the correlation between cosmic strings ($G\mu$) and extra neutrino species ($\neff$) depends on the  data sets used. When relatively low $\ell$ data is used (WMAP7) these two components are anticorrelated; thus, the inclusion of cosmic strings into the model not only did account for the extra radiation species needed, but it actually lowered the need for them. However, when relatively higher $\ell$ are included, the anticorrelation becomes correlation, and an extra cosmic string component asks for a higher contribution from extra relativistic species. This (at first sight) unexpected effect may be understood by noting that a change in $\neff$ changes the CMB temperature anisotropies power spectrum basically in two ways: the height of the first peak and the position of the higher peaks. When only WMAP7 data are taken into account, only the height of the first peak is of importance, and cosmic strings are used to help in this endeavor. However, when higher $\ell$-s are taken into account, 
the position of the higher peaks gains more relative importance, and this is counteracted by a correlation with cosmic strings. This correlation/anticorrelation effect not only happens between $G\mu$ and $\neff$, the same effect can be observed also between $\Omega_b$ and $\neff$. 

When non-CMB data are taken into account, the available parameter space shrinks considerably. This effect comes mostly from the HST value for $h$. Typically, when  the HST value of $h$ is not taken into account in the analysis, rather high values of $h$ are favored, and this is why adding the more stringent prior given by HST the parameters get much more constrained.  The non-CMB data help constrain $\neff$ considerably better when only WMAP7 data  are used, and they break most of the degeneracies when also SPT data are used.  It would be interesting to analyze  whether new CMB data alone, for example Planck data, could 
break the degeneracies between $\neff$ and $G\mu$ without the need of non-CMB external data \cite{lsu}.

The analysis also shows other remarkable results. CMB data alone prefer a high value for the $\neff$, higher than the standard three neutrino species, when strings are present. However, when the HST value of $h$ is included, the results are consistent with only three neutrinos. In most cases, the string contribution is small, and models with no strings are consistent with the data; the case where CMB+SPT are used with non-CMB data are the most permissive with strings, since a $2\sigma$ {\it preference} (or rather, {\it hint}) is obtained.
In many cases an $n_s=1$ perfectly consistent with the data. In fact, we studied what is the goodness of fit for a HZ$+G\mu+\neff$ model, fitting to all the CMB data, and found that the model fits the data extremely well. Actually, the likelihood for HZ$+G\mu+\neff$ is actually very similar to a PL$+G\mu+\neff$, even though the latter has one more parameter. However, the inclusion of non-CMB data will also disfavour this model.

A certain number of cosmic strings are preferred to fit the data, and those cosmic strings would create some gravitational waves that are responsible for part of the $\neff$. We have tried to factor out this contribution, as well as the contribution of the standard three neutrinos, to estimate whether some other source to $\neff$ was still needed, and encoded it in $\neff^*$. A value $\neff^*=0$ would mean that all the players in the model are enough to account for all the $\neff$ needed to fit the data, i.e., the three neutrinos and the cosmic strings present do the job. We find that, even though the errors are rather high, in most cases the strings can account for all the extra radiation component, although  the need for yet another extra relativistic source is not excluded. More importantly, sometimes $\neff^*$ is negative, hinting for some incongruence:  the amount of strings chosen from the fitting is too high.  It could be a good measure to disqualify models as not viable, but unfortunately our 
feeling is that the uncertainties in obtaining $\ngwcs$ are too high to make any kind of claim. One measure for the level of uncertainty could be the big difference in numbers for smaller and bigger cosmic strings loops; but all other uncertainties in the way those estimates are obtained may be much bigger. However, should a revised and more consensual version of the gravitational wave predictions from cosmic strings happen to be as high, the impact that strings have in $\neff$ would definitely have to be included and parameter fitting works including cosmic strings would have to be revisited.

In order to fully exploit the direct link between cosmic strings, gravitational waves and $\neff$, models with $\neff^*=0$ were also studied. These models allow for an extra radiation component, but consider that all the extra radiation come from the cosmic string contribution. Thus, these models have only seven parameters, but include cosmic strings and an extra radiation component. In general terms, these models fit the data rather well (in many cases their goodness of fit is similar to models with one parameter more), and if, as mentioned before, the gravitational radiation from cosmic strings was more under control, these models would be excellent candidates to fit the data well, with fewer parameters.

We should mention the shortcomings and possible improvements for this work. The string power spectra used in this work is the one coming from Abelian Higgs field simulations. There are other approaches to obtain the string spectra, which have roughly the same form, but the details might change the results (though we may anticipate that only slightly). However, a better understanding of the power spectra coming from strings at high $\ell$ would be good.

As already mentioned several times, the main source of uncertainty comes from the production of gravitational waves from cosmic strings. There is intense work in this subject from different groups, using different approaches, and there is no consensus in the community. Unlike in the case of CMB predictions from strings, where different approaches give relatively fairly similar results, the loop production and decay process of strings is still a very open issue. One could guess that the outcomes from Nambu-Goto type approaches or from Abelian-Higgs like approaches will be rather different, since the string density, decay mechanisms and loop density and sizes seem to be rather different.  In this work we have used one of the few works which give a recipe  to translate from cosmic string $G\mu$ to gravitational waves, which in turn we transform into $\neff$, but we believe there are many uncertainties and assumptions that need to be checked an improved. We, therefore, consider this work as a first step into the  analysis of the role of cosmic strings as sources of extra radiation component in the universe,  which demands further understanding of the underlaying cosmic string dynamics. Moreover, it would also be interesting to  perform an analysis including extended cosmological  parameters as in \cite{Joudaki:2012fx}.\\

\section{Acknowledgements}
We are grateful to Neil Bevis, Mark Hindmarsh and Martin Kunz for allowing us to use their  cosmic string spectra.  We thank them and also Andrew Liddle for helpful discussions. J.U. acknowledges financial support from the Basque Government (IT-559-10), the Spanish Ministry (FPA2009-10612), and the Spanish Consolider-Ingenio 2010 Programme CPAN (CSD2007-00042). I.S. holds a PhD FPI fellowship contract from the Spanish Ministry of Economy and Competitiveness and J.L a PhD PIF fellowship from the
EHU/UPV. Numerical calculations were performed using the UK National Cosmology Supercomputer (supported by SGI/Intel, HEFCE, and STFC) and the Andromeda cluster of the University of Geneva.

\bibliography{neffgmu}

\end{document}